\documentclass[letterpaper,titlepage,10pt]{article}
\usepackage[utf8]{inputenc} 
\usepackage{textcomp} 
\usepackage{graphicx}  
\usepackage{flafter}  

\usepackage{amsmath,amssymb}  
\usepackage{bm}  
\usepackage{wasysym}

\usepackage{memhfixc}  
\usepackage{pdfsync}  
\newtheorem{jussi}{Conjecture}
\begin{document}
\title{Governing Dynamics from cause and effect:     
\newline
 -A novel formulation for causality with applications in Quantum Gravity  and Economic Theory}
\author{Jussi Lindgren, jussi.i.lindgren(at)helsinki.fi}
\maketitle
\begin{abstract}
This paper defines an equation for causality. This equation is then combined with the postulates of quantum mechanics and mass-energy equivalence to produce a quantum mechanical telegrapher's equation and to reproduce the Klein-Gordon equation. The incompressible Navier-Stokes equations and dynamic general equilibrium in economics (with an interpretation of a Nash equilibrium) are obtained when the equation of causality refers to itself, i.e. when the cause is its own effect. As it is shown that the Klein-Gordon equation is obtained by Wick rotating the cause vector with de Broglie angular frequency, this paper postulates an equation for Quantum Gravity, which relates the Navier-Stokes equations to the Einstein Field Equations of General Relativity.
\end{abstract}
\begin{quote}

Everything should be made as simple as possible, but not simpler. -Albert Einstein
\end{quote}
\section{Introduction}
In the field of natural sciences, one essentially studies the relations between concepts that are found in nature. These concepts include e.g. the rules that govern the motion of celestial bodies, electricity, biological processes, economic decisions and so forth. Whatever the concepts, the essence of science is based on observation: in the history mankind, humans have observed certain reoccurring patterns of nature and thus these everyday experiences are seen to form the natural laws of nature. One must keep in mind that these observations, for example the rising and setting of the sun are essentially statistical in their nature. The natural laws are perceived as laws because we witness some fairly stable relations between concepts in our everyday lives. In purely logical terms this does not, however, mean that these relations are of causal logical nature. In logical terms there is always a possibility that we do not see the rising sun some misty morning. Therefore we should appreciate the inherently statistical nature of our knowledge of the natural world. Despite of the fact the our knowledge is of statistical nature, the assumed statistics can produce essentially different categories in terms of perception. If one does an empirical study of a given phenomenon and finds that something occurs always within one set, given that there is a fairly large number of observations, one easily deduces that the natural law there is that some variable X that takes aways values in some set Y. This deception is seen easily by considering the normal distribution, because of the exponential decay of the probability density, an observer that would not know of the background of the experiment, would easily say that something happens always in the range of, say 4 standard deviations of the mean. Assuming that the data generating process is indeed normal, this observation would be verified by empirical repetition of the experiment, but alas, would ultimately be wrong. When Isaac Newton proposed his famous laws, he did not derive these laws from anywhere, he just inferred them from the natural experiments he did. The second law of Newton can be stated as
\begin{equation}
\vec{F}=m\vec{a}
\end{equation}
One could always wonder, whether Newton just actually observed
\begin{equation}
\vec{F}=m\vec{a}+\vec{\epsilon}
\end{equation}
where $\epsilon$ is some realization of a random variable, with negligible variance around zero.
This problem described above basically means that we cannot establish the true laws of nature by observation. Therefore one ends up with two options: 
\begin{enumerate}
\item Either one collects all the observational relationships of nature and tries to find a common set of minimal rules that produce the (regressional) equations that we assume nowadays to be the natural laws of nature. This process is inherently empirical and statistical in its nature. If one has a collection of natural laws, let's say the laws of gravitation and electricity, the obvious regressional equation we end up with is the classical laws of Isaac Newton. This is what happened essentially before the twentieth century. 
\item Or guess the right axioms of natural laws that would imply the right equations for the empirist to verify.
\end{enumerate}
For the aesthetically and perhaps mathematically oriented scientist the second option is preferable. We must note that by definition the right axioms of nature must be indeed guessed correctly. One cannot deduce them from some other conditions. Therefore, one can say that for the perfect theory we must have a good guess or either would need an oracle. Ultimately, in terms of epistemology, these two options are unfortunately essentially equivalent. The second option is what I call epistemological regression, as one has to verify the implications of the axioms continuously with the new data that can be observed. In high energy particle physics this is evident; we witness new conditions that are not to be found naturally in the nature. So both mechanism for scientists are equally good or equally bad. There are two important features of this method of scientific discovery
\begin{enumerate}
\item Verification of theories are always statistical in their nature
\item In both approaches what is required is the consistency of relations between different concepts
\end{enumerate}
The present study tries to proceed with axioms and tries to have consistency between different concepts
\section{The functional and causal view of nature}
Consider now a model of physical reality. In simple terms, we try to establish a functional relationship between two objects in space-time. Our space-time is the normal euclidean coordinate space 
\begin{equation}
x,y,z,t \in \mathbb{R}^3\times \mathbb{R}
\end{equation}
Let us now assume that we have a function that maps 
\begin{equation}
\vec{f}(x,y,z,t):x,y,z,t \longrightarrow \mathbb{R}^3
\end{equation}
The fundamental property of any function that is a model of nature (or it's subsets) are the changes in the value of the function as time goes by. After all, the human perception of nature is based on comparison and comparison takes time. Let us now study what happens to this function in time by taking it's Total Derivative with respect to time. 
\begin{equation}
\frac{df_i}{dt}=\frac{\partial f_i}{\partial t}\frac{dt}{dt}+\frac{\partial f_i}{\partial x}\frac{dx}{dt}+\frac{\partial f_i}{\partial y}\frac{dy}{dt}+\frac{\partial f_i}{\partial z}\frac{dz}{dt}
\end{equation}
We can simplify this a bit
\begin{equation}
\frac{df_i}{dt}=\frac{\partial f_i}{\partial t}+\frac{\partial f_i}{\partial x}\frac{dx}{dt}+\frac{\partial f_i}{\partial y}\frac{dy}{dt}+\frac{\partial f_i}{\partial z}\frac{dz}{dt}
\end{equation}
This is the fundamental relation that describes observations in the space-time. We can express it in more convenient notation by making the conventions
\begin{equation}
\dot{x}=\frac{dx}{dt}
\end{equation}
\begin{equation}
\dot{y}=\frac{dy}{dt}
\end{equation}
\begin{equation}
\dot{z}=\frac{dz}{dt}
\end{equation}
\begin{equation}
\dot{X}=(\dot{x},\dot{y},\dot{z})
\end{equation}
\begin{equation}
\nabla f_i=(\frac{\partial f_i}{\partial x},\frac{\partial f_i}{\partial y},\frac{\partial f_i}{\partial z})
\end{equation}
\begin{equation}
\vec{f}=(f_x,f_y,f_z)
\end{equation}
So that we can express the Total Derivative
\begin{equation}
\frac{d\vec{f}}{dt}=\frac{\partial \vec{f}}{\partial t} +\left (
\begin{array}{ccc}
\frac{\partial f_x}{\partial x} & \frac{\partial f_x}{\partial y} & \frac{\partial f_x}{\partial z}\\
\frac{\partial f_y}{\partial x} & \frac{\partial f_y}{\partial y} & \frac{\partial f_y}{\partial z}\\
\frac{\partial f_z}{\partial x} & \frac{\partial f_z}{\partial y} & \frac{\partial f_z}{\partial z}\\\end{array} \right ) \left ( \begin{array}{c}
\dot{x}  \\
\dot{y}  \\
\dot{z}  
\end{array} \right )\end{equation}
Then we can write
\begin{equation}
\frac{d\vec{f}}{dt}=\frac{\partial \vec{f}}{\partial t}+\nabla \vec{f}\dot{\vec{X}}'
\end{equation}
Let us now consider for a while what the Total Derivative of the vector field $\vec{f}$ with respect to time means. First of all the vector field is supposed to model some subset of the physical nature. This means that the Total Derivative with respect to time means that the Total Derivative represents the change in the physical system as a whole. The vector field itself represents a quite comprehensive picture: for each point in the space-time continuum it encodes two properties, the strength of some object and the direction of some object.
\subsection{The cause and effect}
It is natural for us to think that for every effect there is a cause. This perception of the natural world turns to be quite useful, as essentially the identification of cause and effect relationships enables us to build models and therefore enables us the make predictions about the future. After all, we are interested in the future state of things. Cause and effect should be seen in a dualistic context:
\begin{enumerate}
\item When there is a cause, there is an effect.
\item When there is an effect, there is an underlying cause to it.
\item Mathematically causality is equivalent to differential equations involving the Total Derivative with respect to time
\end{enumerate}
These are essentially the most profound axioms of science. Otherwise everything would be beyond human comprehension. For every set $C$ which the principle of causality applies, we can label to be natural. The concepts outside the set $C$ can be labelled as unnatural or theological.
\newline
How to encode these axioms of perception into the context of a differentiable vector field $\vec{f}$? We proceed by identifying the Total Derivative with respect to time as the effect of a cause. Additionally we assume that there is a differentiable vector field $\vec{g}$ that represents the cause. We can therefore form the differential equation
\begin{equation}
\frac{d\vec{f}}{dt}=\vec{g}
\end{equation}
Using the derived results earlier, we write
\newline
\begin{equation}
 \boxed{\frac{\partial \vec{f}}{\partial t}+\nabla \vec{f}\dot{X}'=\vec{g}}
 \end{equation}
 \newline

This is the fundamental \textbf{Equation of Causality}.
\subsection{On the nature of time}
When one thinks of causality, one has to address the question what is required to acknowledge and verify causal relationships. As stated earlier, human perception is based on comparison. We compare things to differentiate classes of objects. Suppose that there were no changes in a physical system. For an external observer of the system it would seem that the system is freezed and time does not elapse. Therefore time does not go forward for an external observer of a system if we have
\begin{equation}
\frac{\partial f}{\partial t}=0
\end{equation}
This implies that one has
\begin{equation}
\nabla \vec{f}\dot{X}'=\vec{g}
\end{equation}
It is extremely important to realize that the cause and effect is still there. In other words the system has time that evolves, but because the cause vector $\vec{g}$ balances the system, the external observer could not distinguish whether time elapses or not. \emph{In other words, time is a subjective observable.}
Note that we have only assumed so far that there is causality in the space-time continuum. The Equation of Causality is rather general as it assumes only that information of a system in space-time continuum can be encoded in differentiable vector fields. 
\subsection{The Equation of Causality for information flow}
Let us now consider the space-time continuum again. Suppose that each point in space-time $\mathbb{R}^3\times \mathbb{R}$ contains hidden information of some sort. At each point in the space-time the cause vector $\vec{g}$ gives some state value to the point $\vec{x},t$.  
\begin{equation}
x,y,z,t \longrightarrow (g_x,g_y,g_z)
\end{equation}
So the vector reads from 4 inputs and returns 3 outputs, in other words \emph{it compresses information}. On the other hand we have the transition of states according to

\begin{equation}
\left (\begin{array}{c}
x \\
y \\
z \\
t \\
\end{array} \right ) \longrightarrow
\left (\begin{array}{c}
x' \\
y'\\
z' \\
t' \\
\end{array} \right ) 
\end{equation}
This transition of states is completely specified by the function
\begin{equation}
\frac{d\vec{X}(t)}{dt}=\dot{\vec{X}}=\vec{v}
\end{equation}
Suppose now that we want to ask: what is the effect of the given cause $\vec{g}$ on the transition of states?

Accordingly the Equation of Causality becomes
\begin{equation}
\frac{\partial \vec{v}}{\partial t}+\vec{v}\cdot \nabla \vec{v}=\vec{g}
\end{equation}

Consider now the following inner product
\begin{equation}
\vec{v}\cdot \vec{g}=\frac{\partial\frac{1}{2} \vec{v}^2}{\partial t}+\vec{v}\cdot (\vec{v}\cdot \nabla \vec{v})
\end{equation}
We will now make of use of the following vector calculus identity
\begin{equation}
\vec{v}\cdot \nabla \vec{v}=\nabla \frac{1}{2}\vec{v}^2-\vec{v}\times (\nabla \times \vec{v})
\end{equation}
Substituting this into equation, one has
\begin{equation}
\vec{v}\cdot \vec{g}=\frac{\partial\frac{1}{2} \vec{v}^2}{\partial t}+\vec{v}\cdot (\nabla \frac{1}{2}\vec{v}^2-\vec{v}\times (\nabla \times \vec{v}))
\end{equation}
Using the notation
\begin{equation}
\vec{L}=\vec{\omega}\times \vec{v}
\end{equation}
where $\vec{\omega}=\nabla \times \vec{v}$ and using the fact that vector product is anti commutative, we can write
\begin{equation}
\vec{v}\cdot \vec{g}=\frac{\partial\frac{1}{2} \vec{v}^2}{\partial t}+\vec{v}\cdot (\nabla \frac{1}{2}\vec{v}^2+\vec{L})
\end{equation}
Using the identities of vector calculus further, we have
\begin{equation}
\vec{v}\cdot \vec{L}=0
\end{equation}
So that in the end one has the \textbf{Master Equation}
\begin{equation}
\boxed{\vec{v}\cdot \vec{g}=\frac{\partial\frac{1}{2} \vec{v}^2}{\partial t}+\vec{v}\cdot \nabla \frac{1}{2}\vec{v}^2}
\end{equation}

\section{Physics from the equation of causality}
So far there has been no physics in this paper, only logical considerations. Let us try to add some physical considerations into our Master Equation

\begin{equation}
\vec{v}\cdot \vec{g}=\frac{\partial\frac{1}{2} \vec{v}^2}{\partial t}+\vec{v}\cdot \nabla \frac{1}{2}\vec{v}^2
\end{equation}

Let us scale the equation with a scale factor $m\in \mathbb{R}$
Then we have
\begin{equation}
m\vec{v}\cdot \vec{g}=\frac{\partial\frac{1}{2} m\vec{v}^2}{\partial t}+\vec{v}\cdot \nabla \frac{1}{2}m\vec{v}^2
\end{equation}
We shall use the notation 
\begin{equation}
E=\frac{1}{2}m\vec{v}^2
\end{equation}
and
\begin{equation}
\vec{p}=m\vec{v}
\end{equation}
Perhaps it is more convenient to rearrange the formula 
\begin{equation}
\frac{\partial E}{\partial t}=\vec{p}\cdot \vec{g}-\vec{v}\cdot \nabla E
\end{equation}

We can also see
\begin{equation}
\frac{\partial E}{\partial t}=\vec{p}\cdot (\vec{g}-\frac{1}{m}\nabla E)
\end{equation}
So that the evolution of $E$ is proportional to the vector difference of the cause vector $\vec{g}$ and the gradient of $E$.
Let us assume that the cause vector is of the form 
\begin{equation}
\vec{g}=\alpha \vec{v} 
\end{equation}

Substituting this into the Master Equation
\begin{equation}
\frac{\partial E}{\partial t}=\vec{p}\cdot (\alpha \vec{v} 
-\frac{1}{m}\nabla E_k)
\end{equation}
We can the rearrange
\begin{equation}
\frac{\partial E}{\partial t}=\alpha m \vec{v}^2 +\vec{p}\cdot (-\frac{1}{m}\nabla E)
\end{equation}
Which is the same as
\begin{equation}
\frac{\partial E}{\partial t}=2\alpha E  +\vec{p}\cdot (-\frac{1}{m}\nabla E)
\end{equation}

\subsection{Enter the physics: Operator Substitution}
Notice now that $E$ can be interpreted as the total energy of the system and $\vec{p}$ can be interpreted as the total linear momentum of the system. We interpret the scale parameter $m$ later on. Let us now make use of the postulates of quantum mechanics and implement the differential operator representation of momentum and total energy as follows
\begin{equation}
E \longrightarrow -\frac{\hbar}{i}\frac{\partial }{\partial t}
\end{equation}
\begin{equation}
\vec{p} \longrightarrow \frac{\hbar}{i}\nabla
\end{equation}
We then have the quantum mechanical version for our Master Equation
\begin{equation}
-\frac{\hbar}{i}\frac{\partial ^2 \psi}{\partial t^2}=-\frac{2\alpha \hbar}{i} \frac{\partial \psi}{\partial t}-\frac{\hbar ^2}{m}\Delta\frac{\partial \psi}{\partial t}
\end{equation}
Let us divide the equation by the factor $i\hbar$ 
\begin{equation}
\frac{\partial ^2 \psi}{\partial t^2}=2\alpha \frac{\partial \psi}{\partial t}-\frac{\hbar }{im}\Delta\frac{\partial \psi}{\partial t} 
\end{equation}
Taking out the time derivative operator under the laplacian and substituting the total energy one has 
\begin{equation}
\frac{\partial ^2 \psi}{\partial t^2}=2\alpha \frac{\partial \psi}{\partial t}+\frac{E}{m}\Delta \psi 
\end{equation}

We take note that $E=\frac{1}{2}mv^2$ is the total energy of the system (free particle). We can then write
\begin{equation}
\frac{\partial ^2 \psi}{\partial t^2}=2\alpha \frac{\partial \psi}{\partial t}+\frac{1}{2}v^2\Delta \psi 
\end{equation}

Rearranging terms, we have

\begin{equation}
\boxed{\frac{1}{v^2}\frac{\partial ^2 \psi}{\partial t^2}-\frac{2\alpha}{v^2} \frac{\partial \psi}{\partial t}=\frac{1}{2}\Delta \psi }
\end{equation}
Which we shall call the Quantum Telegraph Equation (QTE).

Let us now consider the parameter $\alpha$. Let us have
\begin{equation}
\alpha=\frac{1}{2}mv^2\frac{i}{\hbar}=E \frac{i}{\hbar}=i\omega
\end{equation}
where 
\begin{equation}
E=\hbar \omega
\end{equation}

\subsection{Generalized Klein-Gordon equation}
Considering the QTE,
We can use the substitution 
\begin{equation}
\psi=e^{-\alpha t}\Psi
\end{equation}
Then $\Psi$ satisfies the PDE
\begin{equation}
\frac{\partial ^2\Psi}{\partial t^2}=\frac{1}{2}v^2 \Delta \Psi +\alpha^2 \Psi
\end{equation}
or
\begin{equation}
\boxed{\frac{1}{v^2}\frac{\partial ^2\Psi}{\partial t^2}=\frac{1}{2}\Delta \Psi +\frac{\alpha^2}{v^2} \Psi}
\end{equation}
which we shall call the generalized Klein-Gordon equation.

Substituting the parameter $\alpha$ on has
\begin{equation}
\boxed{\frac{1}{v^2}\frac{\partial ^2\Psi}{\partial t^2}=\frac{1}{2}\Delta \Psi -\frac{1}{4}\frac{m^2v^2}{\hbar^2} \Psi}
\end{equation}
Let us multiply both sides with $-\frac{\hbar^2}{m}$
\begin{equation}
\boxed{-\frac{\hbar^2}{mv^2}\frac{\partial ^2\Psi}{\partial t^2}=-\frac{\hbar^2}{2m}\Delta \Psi +\frac{1}{4}mv^2 \Psi}
\end{equation}
using again
\begin{equation}
v^2=c^2
\end{equation}
one has
\begin{equation}
\boxed{-\frac{\hbar^2}{E_0}\frac{\partial ^2\Psi}{\partial t^2}=-\frac{\hbar^2}{2m}\Delta \Psi +\frac{1}{4}E_0\Psi}
\end{equation}
\subsection{Generalized Schrodinger equation}

Substituting the parameter $\alpha$, one has

\begin{equation}
\boxed{\frac{1}{v^2}\frac{\partial ^2 \psi}{\partial t^2}-\frac{im}{\hbar} \frac{\partial \psi}{\partial t}=\frac{1}{2}\Delta \psi }
\end{equation}

Or

\begin{equation}
\boxed{\frac{-\hbar^2}{mv^2}\frac{\partial ^2 \psi}{\partial t^2}+i\hbar \frac{\partial \psi}{\partial t}=-\frac{\hbar^2}{2m}\Delta \psi }
\end{equation}

We can immediately obtain the Schrodinger equation by taking the limit
\begin{equation}
\lim_{ v^2 \to \infty}
\end{equation}
On the other hand, if we set that
\begin{equation}
v^2=c^2
\end{equation}
we will have
\begin{equation}
\boxed{\frac{-\hbar^2}{E_0}\frac{\partial ^2 \psi}{\partial t^2}+i\hbar \frac{\partial \psi}{\partial t}=-\frac{\hbar^2}{2m}\Delta \psi }
\end{equation}
where $E_0=mc^2$ is the rest energy of the mass $m$. We can see from the equation that for all but very tiny particles in terms of rest mass the quantum equation is the Schrodinger equation, so 
\begin{equation}
\frac{-\hbar^2}{E_0}\frac{\partial ^2 \psi}{\partial t^2} \sim 0
\end{equation}

\subsection{Towards Quantum Gravity}

Recently many researchers have studied the connection between the incompressible Navier-Stokes equations and the Einstein field equations of General Relativity. It has been shown that there are indeed deep mathematical connections between the two. Especially interesting are the very recent works of Bredberg, Keeler, Lysov and Strominger, see e.g. \cite{harvard}. Another interesting result is due to Padmanabhan, see \cite{pad}. Therefore the suggested way forward would to try to establish some sort of equivalence between the Navier-Stokes equations and gravitation.
We proceed by use of a lemma regarding the Navier-Stokes equations. Regarding the regularity of Navier-Stokes equations, see \cite{jussi}

Consider now again the Equation of Causality
\begin{equation}
\frac{\partial \vec{v}}{\partial t}+\vec{v}\cdot \nabla \vec{v}=\vec{g}
\end{equation}
We now again turn to self-referential question what is the effect of a cause to itself?
We then have
\begin{equation}
\frac{\partial \vec{v}}{\partial t}+\vec{v}\cdot \nabla \vec{v}=\vec{v}
\end{equation}
Let us now use the Helmholtz decomposition theorem of vector fields and decompose the vector field $\vec{v}$ into 
\begin{equation}
\vec{v}=-\nabla p + \nabla \times \vec{A}
\end{equation}
One then has the following equation 
\begin{equation}
\frac{\partial \vec{v}}{\partial t}+\vec{v}\cdot \nabla \vec{v}=-\nabla p + \nabla \times \vec{A}
\end{equation}
With the condition that
\begin{equation}
\nabla \cdot \vec{A}=0
\end{equation}
On the other hand we have the vector calculus identity
\begin{equation}
\nabla \times (\nabla \times \vec{F})=\nabla(\nabla \cdot \vec{F})-\Delta \vec{F}
\end{equation}
Suppose now that the vector field $\vec{A}$ is derived as follows
\begin{equation}
\vec{A}=-\nabla \times \vec{v}
\end{equation}
Then we have
\begin{equation}
\nabla \times \vec{A}=-\nabla \times \nabla \times \vec{v}
\end{equation}
and thus, noting that divergence of $\vec{A}$ is zero
\begin{equation}
\nabla \times \vec{A}=\Delta \vec{v}
\end{equation}
Therefore we obtain
\begin{equation}
\frac{\partial \vec{v}}{\partial t}+\vec{v}\cdot \nabla \vec{v}=-\nabla p + \Delta \vec{v}
\end{equation}
with 
\begin{equation}
\nabla \cdot \vec{v}=0
\end{equation}
This is a magnificent result.The two following equations are equivalent
\begin{equation}
\frac{\partial \vec{v}}{\partial t}+\vec{v}\cdot \nabla \vec{v}=\vec{v}
\end{equation}
\begin{equation}
\frac{\partial \vec{v}}{\partial t}+\vec{v}\cdot \nabla \vec{v}=-\nabla p + \Delta \vec{v}
\end{equation}
with the condition
\begin{equation}
\nabla \cdot \vec{v}=0
\end{equation}
and some scalar field $p$ and assuming that the vector field $\vec{v}$ is square-integrable. Note that $p$ must obey
\begin{equation}
\Delta p =0
\end{equation}

When we derived the Quantum Telegraph Equation and the generalized Klein-Gordon equation, essentially we quantized the system
\begin{equation}
\frac{\partial \vec{v}}{\partial t}+\vec{v}\cdot \nabla \vec{v}=\alpha\vec{v}
\end{equation}
Substituting alpha we have
\begin{equation}
\frac{\partial \vec{v}}{\partial t}+\vec{v}\cdot \nabla \vec{v}=\frac{i}{\hbar}\frac{1}{2}{mc^2} \left (-\nabla p +\Delta \vec{v}\right )
\end{equation}
So we have in a sense "Wick rotated Navier-Stokes equations". 
Suppose now we operate the above equation with a partial time derivative operator
\begin{equation}
\frac{\partial }{\partial t}\left (\frac{\partial \vec{v}}{\partial t}+\vec{v}\cdot \nabla \vec{v}\right )=\frac{i}{\hbar}\frac{\partial }{\partial t}\left (\frac{1}{2} mc^2\left (-\nabla p +\Delta \vec{v}\right )\right )
\end{equation}

Let us do the inverse operator substitution according to the postulate of quantum mechanics

\begin{equation}
E_0 \longrightarrow -\frac{\hbar}{i}\frac{\partial }{\partial t}
\end{equation}
So that we have

\begin{equation}
\frac{\partial }{\partial t}\left ( \frac{\partial \vec{v}}{\partial t}+\vec{v}\cdot \nabla \vec{v} \right )=-E_0\left (\frac{1}{2}mc^2\left (-\nabla p +\Delta \vec{v}\right )\right )
\end{equation}

Let us assume that the energy term $E_0$ is equal to some rest energy
\begin{equation}
E_0=m_0c^2
\end{equation}

\begin{equation}
\boxed{\frac{\partial }{\partial t}\left ( \frac{\partial \vec{v}}{\partial t}+\vec{v}\cdot \nabla \vec{v} \right )=-\frac{1}{2}m_0 mc^4\left (-\nabla p +\Delta \vec{v}\right )}
\end{equation}

Einstein Field Equations are given by
\begin{equation}
R_{\mu \nu}-\frac{1}{2}g_{\mu \nu}R=\frac{8\pi G}{c^4}T_{\mu \nu}
\end{equation}
where $R_{\mu \nu}$ is the Ricci curvature tensor, $g_{\mu \nu}$ is the metric tensor, $R$ is scalar curvature, $G$ is Newton's gravitational constant, $c$ is the speed of light and $T_{\mu \nu}$ is the stress-energy tensor.
Einstein field equations can be cast also in a compact form
Or in amore compact notation
\begin{equation}
G_{\mu \nu}=\frac{8\pi G}{c^4}T_{\mu \nu}
\end{equation}
where $G_{\mu \nu}$ is the Einstein tensor.
\begin{jussi}
The Einstein Field Equations are (at least in some weak sense) equivalent to the equations
\begin{equation}
\boxed{\frac{\partial }{\partial t}\left ( \frac{\partial \vec{v}}{\partial t}+\vec{v}\cdot \nabla \vec{v} \right )=-\frac{1}{2}m_0 mc^4\left (-\nabla p +\Delta \vec{v}\right )}
\end{equation} 
\end{jussi}
There is an analogy to the equations of General Relativity as the right side of the equation combines energy and the stress tensor of the Navier-Stokes equations. We can therefore call it the Stress-Energy tensor.
It should be note that the Nobel laureate John Nash has similar thoughts and he has derived an equation which has a peculiar structure, see \cite{nash}.

Note on the scalar field $p$.
\begin{equation}
\nabla \cdot \vec{v}=0
\end{equation}
Note that the incompressibility requirement states then that
\begin{equation}
\Delta p=0
\end{equation}
So that $p$ is a harmonic scalar field , it could be example of the form
\begin{equation}
p = \frac{1}{r}
\end{equation}
where
\begin{equation}
r^2=x^2+y^2+z^2
\end{equation}
which has an obvious link to Newtonian potentials in physics. Note now that the right side of our equation resembles Newton's universal law of gravitation.

\section{Dynamic General Equilibrium in economic theory}
\begin{quote}
Equations are more important to me, because politics is for the present, but an equation is something for eternity. -Albert Einstein
\end{quote}

In the year 2000, the Fields Medalist Stephen Smale listed 18 top mathematical problems for this century in his article \cite{smale}. One of the problems concerns the general equilibrium theory in mathematical economics. More specifically, problem 8 in the list states that:
\newline
\newline
\emph{Extend the mathematical model of general equilibrium theory to include price adjustments}
\newline

General equilibrium theory is well established since the works of the Nobel laureates Kenneth Arrow and Gerard Debreu \cite{debreu} in the 1950's. For noneconomic readers the general equilibrium refers to existence of fixed points in an economic model where there are many commodities in  a pure exchange economy. The general equilibrium is then the clearing of markets, that is equating the supply and demand as a function of the relative prices of the commodities.  A comprehensive exposition can be found from \cite{MWG}. The theory of general equilibrium is essentially a static theory and it does not describe the dynamics of the price adjustments. Nobel laureate Paul Samuelson however introduced the concept of price tatonnement in his opus \cite{Samuelson}. According to Samuelson, the price adjustment equation is for good $l$
\begin{equation}
\frac{dp_l}{dt}=c_l z_l(p)
\end{equation}
where $p$ refers to prices and $z$ is the excess demand function 
\begin{equation}
z_l(p)=D(p)-S(p)
\end{equation}
and $c_l>0$ is some constant. Stephen Smale himself considered this problem in his article \cite{smale2}.
\newline
Let us now try to formulate this general equilibrium concept in the context of the Equation of Causality. Suppose that we understand the excess the demand to be the fundamental cause to some changes in the economic system. We then ask ourselves: 
\newline

\emph{What is the effect of excess demand on itself?}
\newline

This is a simple and legitimate question, as the excess demand moves prices and prices move excess demand. One should acknowledge the self-referential nature of the question, though. In terms of the equation of causality, we can proceed straightforwardly to encode this question into the following equation
\begin{equation}
\frac{d\vec{z}(\vec{p},t)}{dt}=\vec{z}(\vec{p},t)
\end{equation}
Where the left side of the equation is again the Total Derivative. Although we could assume as many good as we wish, assume now that we have only three goods with prices $\vec{p}(t)=(p_1(t),p_2(t),p_3(t))$. Using the earlier results in this paper, the cause and effect can be then expressed as
\begin{equation}
\frac{\partial \vec{z}(\vec{p},t)}{\partial t}+\nabla \vec{z}(\vec{p},t) \cdot \dot{\vec{p}}=\vec{z}(\vec{p},t)
\end{equation}
This is the most general formulation for the dynamics of excess demand, note that it is a vector field. We now want to take use of Samuelson's tatonnement process and assume
\begin{equation}
\dot{\vec{p}}=\vec{z}(\vec{p},t)
\end{equation}
Substituting this back to the Equation of Causality, one has
\begin{equation}
\frac{\partial \vec{z}(\vec{p},t)}{\partial t}+\nabla \vec{z}(\vec{p},t) \cdot \vec{z}(\vec{p},t)=\vec{z}(\vec{p},t)
\end{equation}
The general equilibrium is then described by a price vector $\vec{p}^*$ such that
\begin{equation}
\vec{z}(\vec{p}^*)=0
\end{equation}

Consider a dynamic game where there are two players. Player 1 has a reaction function 
\begin{equation}
\vec{S}(\vec{D})
\end{equation}
and Player 2 has a reaction function
\begin{equation}
\vec{D}(\vec{S})
\end{equation}
The reaction functions are strategies that are the best responses given the other players strategies. Assuming that we have a Nash equilibrium, we must have
\begin{equation}
\vec{S}(\vec{D(\vec{p})})=\vec{D}(\vec{S(\vec{p})})
\end{equation}
Noting that
\begin{equation}
z=D-S
\end{equation}

So that a general equilibrium corresponds to a Nash equilibrium of a game with best response functions $S$ and $D$.
Let us now study the divergence of the excess demand field
\begin{equation}
\nabla \cdot \vec{z}=\frac{\partial z_1}{\partial p_1}+\frac{\partial z_2}{\partial p_2}+\frac{\partial z_3}{\partial p_3}
\end{equation}
If all goods are normal, we must have negative divergence always. Suppose anyhow that the third good models labour. So we have an economic model where there are two normal consumption goods and also supply and demand of labour. Let us denote the labour by $z_3$. then let us define
\begin{equation}
p_3=\frac{1}{\omega}
\end{equation}
where $\omega$ is the wage-level of labour. Then we have the behavior
\begin{equation}
\frac{\partial z_3}{\partial p_3}>0
\end{equation}
Let us furthermore assume that we have
\begin{equation}
\frac{\partial z_3}{\partial p_3}=-\frac{\partial z_1}{\partial p_1}-\frac{\partial z_2}{\partial p_2}
\end{equation}
with some excess demand functions for labour, good 1 and good 2.
So that for the excess demand we have the equations
\begin{equation}
\frac{\partial \vec{z}(\vec{p},t)}{\partial t}+\nabla \vec{z}(\vec{p},t) \cdot \vec{z}(\vec{p},t)=\vec{z}(\vec{p},t)
\end{equation}
with 
\begin{equation}
\nabla \cdot \vec{z}=0
\end{equation}
From the results for the incompressible Navier-Stokes equations, we have that the dynamic process of general equilibrium can be then described as
\begin{equation}
\frac{\partial \vec{z}(\vec{p},t)}{\partial t}+\nabla \vec{z}(\vec{p},t) \cdot \vec{z}(\vec{p},t)=-\nabla q+\Delta \vec{z}
\end{equation}
with 
\begin{equation}
\nabla \cdot \vec{z}=0
\end{equation}
which are exactly the Navier-Stokes equations! Therefore there is reason to believe that an economy behaves like a fluid in motion. 
\section{Conclusions and further research}
This paper has shown that the laws of physics and economics can be derived from a simple principle of cause and effect. Moreover, we get the relativistic quantum mechanics by applying simply the postulates of operator substitution and mass-energy equivalence. It is conjectured that the time-derivative-energy augmented Navier-Stokes equations provide a correct description of gravitation and thus provides an unified theory of the natural world. In addition, we postulate that electromagnetism can be deduced from the model as the telegraphers equation and Maxwells equation can be linked to each other through the electric field. Obviously the next step is to try to prove the conjectures by showing formally the connection between the augmented Navier-Stokes equations and the Einstein Field equations. In terms of economic theory, it is shown that with two goods and a labour market the dynamic general equilibrium can be modeled as Navier-Stokes equations as well. This is extremely interesting as now we can simulate the excess demand adjustment process by utilizing the well-established numerical schemes for the navier-Stokes system. One should also pursue to show the link between our augmented Navier-Stokes equations and Maxwells equations. For what its worth, the postulates of quantum mechanics should be seen in a light where the energy is generated by the product of a complex function and its complex conjugate.

\end{document}